\documentclass[final,1p,times,nopreprintline]{elsarticle}

\usepackage{amssymb}
\usepackage{lineno}
\usepackage[dvipsnames]{xcolor}
\usepackage[newfloat,frozencache,cachedir=minted-cache]{minted}
\usepackage{caption}
\usepackage{url}

\newenvironment{code}{\captionsetup{type=listing}}{}
\SetupFloatingEnvironment{listing}{name=Listing}

\journal{Computational Materials Science}

\begin{document}

\begin{frontmatter}


\title{mkite: A distributed computing platform for high-throughput materials simulations}

\author[inst1]{Daniel Schwalbe-Koda}
\ead{dskoda@llnl.gov}
\affiliation[inst1]{organization={Lawrence Livermore National Laboratory},
            addressline={7000 East Ave}, 
            city={Livermore},
            postcode={94550}, 
            state={CA},
            country={United States}}

\begin{abstract}
    Advances in high-throughput simulation (HTS) software enabled computational databases and big data to become common resources in materials science.
    However, while computational power is increasingly larger, software packages orchestrating complex workflows in heterogeneous environments are scarce.
    This paper introduces mkite, a Python package for performing HTS in distributed computing environments.
    The mkite toolkit is built with the server-client pattern, decoupling production databases from client runners.
    When used in combination with message brokers, mkite enables any available client to perform calculations without prior hardware specification on the server side.
    Furthermore, the software enables the creation of complex workflows with multiple inputs and branches, facilitating the exploration of combinatorial chemical spaces.
    Software design principles are discussed in detail, highlighting the usefulness of decoupling simulations and data management tasks to diversify simulation environments.
    To exemplify how mkite handles simulation workflows of combinatorial systems, case studies on zeolite synthesis and surface catalyst discovery are provided.
    Finally, key differences with other atomistic simulation workflows are outlined.
    The mkite suite can enable HTS in distributed computing environments, simplifying workflows with heterogeneous hardware and software, and helping deployment of calculations at scale.
\end{abstract}

%


\end{frontmatter}


\section{Introduction}
\label{sec:intro}

In the past two decades, advances in high-performance computing (HPC) expedited data-driven approaches to materials design and discovery \citep{NSTC2011Materials}.
With several materials databases dating back to the early 2000s \citep{Klintenberg2002}, millions of materials properties have been calculated using quantum chemistry, density functional theory (DFT), and many other levels of theory. 
Generalist materials databases such as the Materials Project \citep{Jain2013Commentary}, AFLOW \citep{Curtarolo2012AFLOWLIB.ORG,Calderon2015}, OQMD \citep{Saal2013Materials}, or JARVIS-DFT \citep{Choudhary2020TheDesign} offer a variety of properties curated from DFT calculations, whereas materials data repositories such as NOMAD \citep{Draxl2018}, Materials Cloud \citep{Talirz2020}, or Materials Data Facility \citep{Blaiszik2016,Blaiszik2019a} enable independent calculations to be uploaded and shared with the broader community. 
Several other datasets have been developed independently by academic and industrial groups, facilitating the discovery of many classes of materials using computation \citep{Greeley2006Computational,Setyawan2011High,Armiento2011Screening,Hachmann2011,Aykol2016high,Yan2017Solar,Haastrup2018TheCrystals,Tran2018ActiveEvolution,Winther2019,Boyd2019data,Chanussot2021,Peng2022human}.

Foundational to all these efforts is the development of high-throughput simulation (HTS) software. 
As the number of systems under study increases, managing simulation workflows quickly becomes a bottleneck in computational materials research. 
In particular, the combination of diverse calculations, materials systems, software packages, and hardware architectures hinders the deployment of simulations at scale. 
To solve this problem, several HTS and workflow management software packages were released in the past decade, such as the AFLOW suite \citep{Curtarolo2012AFLOWLIB.ORG,Calderon2015}, AiiDA \citep{Pizzi2016,Huber2020AiiDAProvenance,Uhrin2021WorkflowsWorkflows}, FireWorks \citep{Jain2015Fireworks:Applications}, atomate \citep{Mathew2017Atomate:Workflows}, pyiron \citep{Janssen2019Pyiron:Science}, ASE \citep{HjorthLarsen2017}, JARVIS-Tools \citep{Choudhary2020TheDesign}, signac \citep{Adorf2018SimpleFramework}, Merlin \citep{Peterson2019}, and many others \citep{Zapata2019QMflows:Chemistry,dAvezac2012Genetic,Ioannidis2016molSimplify,Armiento2020Database-DrivenPhysics,Youn2020AMP2:Materials,Tran2018DynamicScience,Rosen2019IdentifyingTheory}.
Tools interfacing other packages have also been created to accelerate domain-specific calculations, including the MAterials Simulation Toolkit (MAST) \citep{Mayeshiba2017TheDiffusion} or the MPinterfaces \citep{Mathew2016MPInterfaces:Systems}.

Despite numerous HTS packages available for materials simulation, several challenges in managing large-scale simulations remain. Common sources of computational overhead in managing HTS simulations include:

\begin{enumerate}
    \item \textit{Adapting job submission/parsing protocols in  diverse hardware architectures or operating system}.
    Whenever the computational resources are heterogeneous, adapting submission engines and scripts may require pre-specifying a target cluster for the job (e.g., as in AiiDA) or using environmental variables to unify job submission tasks in different clusters (e.g., as in FireWorks).
    \item \textit{Connecting to production databases from client workers}.
    Although many workflow management systems rely on databases to store and retrieve job information, adding data directly to databases requires either hosting a local server in production HPC systems or remotely connecting to an external database.
    The drawback of the latter is exposing information from the production database to a client worker, which may be undesirable in distributed computing platforms.
    Furthermore, if production HPC nodes are not connected to the internet, job retrieval from the database may not be possible at runtime.
    \item \textit{Mismatches between research timelines and computational workflows}.
    Whereas HTS workflows are carefully designed prior to deployment, computational materials scientists often prototype results with a variety of methods prior to production runs.
    Furthermore, different stages of the same project may require experimenting with different workflows (e.g., pre-screening with lower levels of theory, further refinement with DFT etc.) or adding dynamic behavior (e.g., active learning). 
    As such, documenting different experiments in a ``computational lab notebook'' requires careful version control with the data, but also flexibility for further prototyping/data analysis.
    \item \textit{Handling and creating jobs with multiple inputs}.
    Using more than one material/molecule as an input adds overhead to workflow managers.
    For example, crystal-molecule interfaces require calculation graphs of both molecular and solid systems prior to obtaining interfacial properties.
    In some cases, the number of inputs is variable and depends on the conformational diversity of molecular systems.
    As such, the complexity of calculations grows for each branch of the calculation graph, where several steps have to be completed before proceeding with downstream tasks.
    \item \textit{Orchestrating jobs from more than one software package when performing calculations}.
    Although several HTS packages can perform such orchestration, distributing different parts of a workflow to diverse, heterogenous hardware is challenging.
    For example, molecular conformers can be generated with less computational resources, and may be effectively deployed in idle local workstations, whereas DFT simulations may require dedicated HPC clusters.
    Manually controlling the distribution of tasks across different computing environments is labor-intensive, but can be automated within client-server patterns.
\end{enumerate}

To tackle these issues, this paper introduces mkite, a software toolkit for distributed high-throughput materials simulations. 
The mkite package decouples data storage, job creation and submission, and task management in heterogeneous computing environments, approaching a distributed computing scheme. 
In this platform, the production database is agnostic to the available computing platforms, and the client in HPC systems does not control the production database, following a ``client-server'' design pattern. 
Although this requires an intermediate layer of abstraction connecting the databases and the job runners such as a message broker, it facilitates scaling by preventing concurrency issues in production databases.
The approach also reduces privacy concerns in distributed computing, as clients only receive information about the jobs to be performed and the production database is better protected from external attacks.
The mkite toolkit also allows researchers to input data from a variety of sources, thanks to a flexible, extensible database schema and serialization. 
Finally, mkite recipes and workflows enable prototyping new computational tasks with any external software while preserving the advantages of recording a data provenance. 
The platform may facilitate simulations, scaling, and visualization of materials in a diversity of data and hardware sources, complementing similar efforts in high-throughput materials simulation.

\section{Architectural Overview}
\label{sec:architecture}

\begin{figure}
    \centering
    \includegraphics[width=\linewidth]{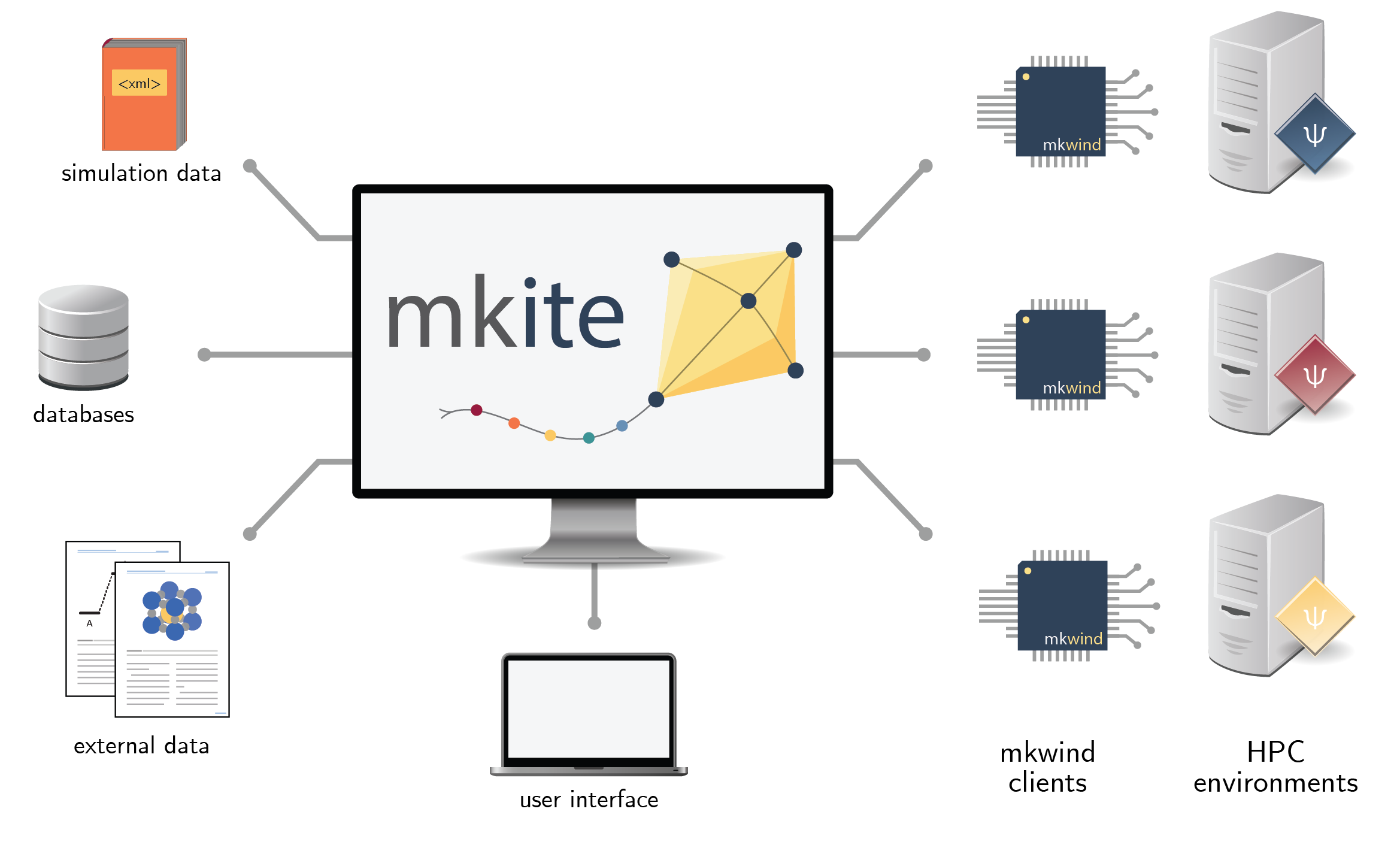}
    \caption{Overview of mkite toolkit. The software provides tools to unify data sources, manage databases, create a user interface, and submit calculations to heterogeneous computing environments.}
    \label{fig:overview}
\end{figure}

An overview of the mkite framework is shown in Fig. \ref{fig:overview}. 
Data is acquired from a variety of sources, and added to the database through user-specific commands or serialization. 
In the current implementation of mkite, the user can access the data through a Django server, which is the object relational mapper (ORM) for a PostgreSQL database (adapted for Python via psycopg2). 
The Django server also allows the user to query, parse, modify, and visualize the data using custom-made views. 
The integration of mkite and Django is further facilitated through serialization from Django Rest Framework (DRF). 
This flexible data structure is also used to standardize communication with the distributed computing software, which is deployed remotely to manage the calculations and simulations.

The mkite interface enables managing different databases by specifying distinct configuration files read by Django. 
As database information should be independent from execution  scripts, each user can use different database credentials for the server (setup tutorials are provided in the mkite documentation).
Then, after a centralized server is set up, simulations in mkite are deployed in a distributed manner. 
This is performed by sending task information to clients --- implemented as the package mkwind --- through a queue system. 
This approach is similar to FireWorks and AiiDA, but instead of customizing the job execution at the server side, jobs are custom-built at the client side using user-defined settings. 
Therefore, the local client is responsible for accessing a task queue, building jobs on local computing environments conditioned to the local configurations, and managing their execution without requiring access to the production database. 
Decoupling functions related to task execution facilitates the scaling and distribution of the jobs, as the main server becomes agnostic to which worker will execute the calculations and the clients do not overwhelm a production database with concurrent requests.

\begin{figure}
    \centering
    \includegraphics[width=.8\linewidth]{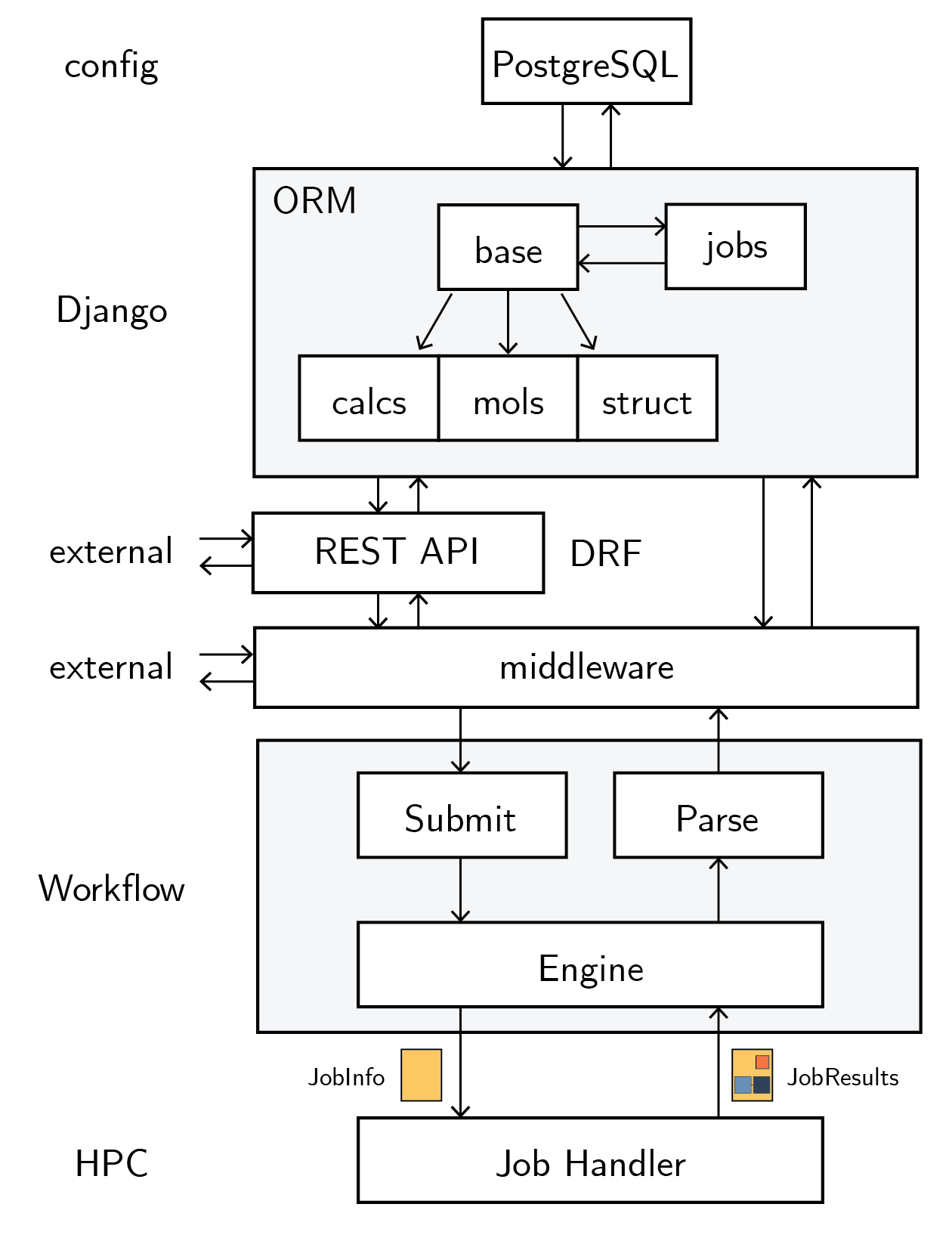}
    \caption{Overview of the mkite substructures. The Django ORM interfaces the PostgreSQL database by defining tables/models, whereas the Django REST Framework and internal middleware provide communication to external packages or internal workflow management subparts.}
    \label{fig:architecture}
\end{figure}

The communication between the mkite substructures is shown in Fig. \ref{fig:architecture}. 
The PostgreSQL database is interfaced by the Django ORM, which in turn uses DRF and the mkite API to communicate with external software and/or internal workflow management commands. 
The REST framework also helps creating Django views for the models, while the middleware enables internal communication between mkite modules without requiring access to the ORM objects.
For example, the \texttt{workflow} module enables: submitting jobs without connecting to a database and managing workflow configurations and definitions. 
This additional layer of abstraction allows external packages to create structured, database-friendly data without access or knowledge of the production database.

In the sections below, this architecture is discussed in detail, along with the design principles that guided its creation. 
In particular, following the logic of Fig. \ref{fig:architecture}, this article is structured as following:

\begin{itemize}
    \item Section \ref{sec:db} discusses the database configuration and schema, and how unstructured data can be added to the database while keeping a structured workflow data.
    \item Section \ref{sec:workflow} discusses the approach to creating jobs and workflows, and their automation using scripts and text files.
    \item Section \ref{sec:exec} showcases the behavior of mkwind, the package that executes the simulations in client hardware. The preparation, submission, and control of jobs is exemplified with multiple queuing systems in distributed platforms.
    \item Section \ref{sec:plugins} explains how the behavior of mkite can be scaled with configuration files and customized with other computational tasks. Different job recipes can be encoded in different packages, decoupling their development and providing flexibility for unique projects.
    \item Section \ref{sec:examples} provides examples of mkite's usage in data analysis and combinatorial calculations. The workflow and simulation steps are discussed in detail for two case studies as examples to the reader.
    \item Section \ref{sec:disc} compares mkite with other HTS software and discusses their complementary behavior. Limitations and future directions of mkite are also highlighted.
\end{itemize}

\section{Database}
\label{sec:db}

The integration between mkite and PostgreSQL enables fast, scalable queries while structuring a schema for complex workflows. 
Although these approaches are also possible with a NoSQL database such as MongoDB, PostgreSQL enables SQL \texttt{JOIN} operations between different tables, which can be useful for combining heterogeneous data sources in complex workflows or connecting simulations from different branches of a calculation graph (e.g., supramolecular interactions). 
The seamless integration between Django and PostgreSQL also facilitates data management, visualization, and analysis beyond SQL-based queries. 
In particular, Django enables the visualization of large amounts of data from different sources, which is often useful in combinatorial spaces of materials simulations \citep{Schwalbe-Koda2021a}.

To describe materials simulation workflows, the mkite database is structured as a graph with three types of nodes: Jobs, ChemNodes, and CalcNodes. 
Jobs contain instructions for performing calculations. 
ChemNodes are generic structures for materials, including crystals, molecules, interfaces, and more. 
Finally, CalcNodes represent physical attributes from ChemNodes deriving from calculations, including energies, forces, density of states (DOS), featurization etc.
Below, these three types of nodes are described in detail, along with their functions.

\subsection{Job schema}
\label{sec:db:job}

\begin{figure}
    \centering
    \includegraphics[width=\linewidth]{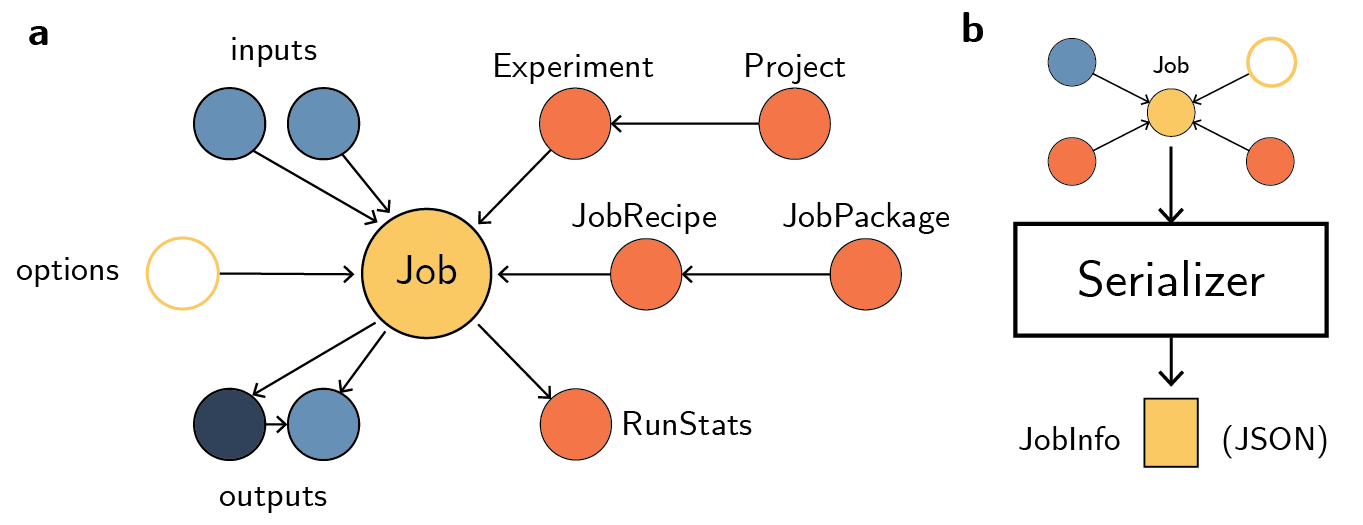}
    \caption{\textbf{a}, Database relations for Job table. A Job has relationships with input nodes (many-to-many), job metadata (foreign key), job execution options (JSON field), and job execution statistics (RunStats, one-to-one). Jobs can produce ChemNodes and CalcNodes as outputs (many-to-many fields). \textbf{b}, Schematic of the serialization of Job nodes. Each Job is joined with its corresponding metadata and inputs, and the database objects are given to a serializer implemented with DRF. This serializer produces a dictionary-like output that can be saved as a JSON file.}
    \label{fig:job-model}
\end{figure}

Central to the mkite code is the Job model/table. 
Figure \ref{fig:job-model}a exemplifies how the Job table relates to other tables. 
The Job table connects metadata on what computational task will be executed (JobRecipe), the inputs for the task, and the outputs generated from the task completion. 
In general, a JobRecipe has only information on the name of the task to be executed and the package (JobPackage) to be called. 
The options to run this job, such as simulation parameters, are provided in the job itself through a JSON field, or assumed as the default options implemented in the recipe.
Input structures to be simulated are provided separately, through a many-to-many relationship between ChemNodes and Jobs in the database.
In addition to the information relevant for execution, a Job contains custom metadata, such as tags or experiments, that allow users to retrieve jobs relevant to a particular project.
While inputs, experiments, and recipes are used to create a job object in the database, each job can create outputs and models containing metadata on the job execution.

Jobs tables aggregate information on the task to be executed. 
However, in distributed computing environments, it is undesirable for a computing resource to access the production database to retrieve the task at hand. 
Although this coupling enables reading/writing information directly from/to the database, an advantage explored by packages such as AiiDA or FireWorks, it may create concurrency problems at large throughputs. 
Another problem with this approach is the requirement of connection to the database itself, which may expose access credentials to the executing platform or impose the requirement of internet connection in production nodes. 
To bypass these problems, database models at mkite can be (de)serialized using DRF, and streamed directly from an API or a message broker. 
Figure \ref{fig:job-model}b shows an example of a Job that is converted to a JSON file that contains all the information of the job, such as serialized identifiers, inputs, recipe, and others. 
Despite the additional overhead in (de)serializing the database models compared to direct access, the REST framework provides the following benefits: 
(1) the structured data can be transferred to other computers via messaging protocols without exposing the database; 
(2) jobs can produce structured data by following the (de)serializer convention, thus bypassing the need for a database connection;
and (3) updating/adding information to the database can be performed using a \texttt{PUT}/\texttt{POST} operation in a REST API on top of the Django server, which includes validation and is an ACID transaction in PostgreSQL.

The serialization layer allows adding job outputs to the database after the task execution. 
Although the remote clients do not have access to the database, they can create serialized files containing the job identifier plus the results of the calculation. 
The serialized file can be easily parsed into the database using the REST framework or a Django command.
Thus, after each calculation is performed, its results are sent to another message queue that is monitored by the production database.
Along with the output ChemNodes and CalcNodes, results often contain metadata on the job, including where it was executed, which resources it used, its duration and any other user-specified information.

A consequence of the (de)serialization is the need to deal with mutable dictionaries in Python. 
To facilitate handling these objects in Python, additional middleware is created to impose a schema on the files (see Fig. \ref{fig:architecture}). 
These models, based on the package \texttt{msgspec} in the current implementation, can quickly read/write files to disk, validate data types, and extend functionalities as Python objects.
In this implementation, the class containing all the outbound data from a Job is called JobInfo, whereas the serialized data to be added to the database is called JobResults. 
Both contain serialized data from other nodes, including ChemNode and CalcNode (discussed below).

\subsection{Node schema}
\label{sec:db:node}

Whereas jobs are intrinsically well-structured, schemas for materials and molecules may exhibit a variety of representations.
This diversity of data formats has promoted the use of NoSQL databases such as MongoDB in the community.
Despite the flexibility of NoSQL approaches for materials simulations, relational databases can effectively perform \texttt{JOIN} operations using SQL, which is useful when selecting database entries matching certain criteria. 
From the perspective of a computational practitioner, this structured data may also help keeping track of all calculations and experiments performed over the course of a given project, approaching an ``electronic lab noteboook'' for computational science. 
At the same time, however, evolving database schemas can be difficult to handle and broadcast across projects with different needs. 
To comply with these conflicting needs, mkite stores chemical data as ChemNodes --- which contains the information about the atomistic structure of the chemical/materials system, including its chemical composition, atomic numbers etc. --- and property data as CalcNodes --- which store information of energies, forces, band structures etc.
ChemNodes and CalcNodes are designed as separate tables in mkite, with a CalcNode requiring a ChemNode as a foreign key, assuming properties are defined for one given structure.

In mkite, two options are available to address the model diversity in ChemNodes: (1) storing all relevant data in flexible JSON fields; or (2) creating tables for storing the structured information of interest. 
In the former, a hybrid NoSQL-relational database is created, i.e. information on jobs and workflows is stored in well-structured tables, but information about the chemical system is stored in a JSON field. 
This approach enables users to avoid developing schemas, as long as user-defined job inputs/outputs are kept consistent throughout the mkite workflows. 
This approach also integrates well with existing software packages such as pymatgen or ASE, where the chemical data is easily to/from dictionaries and other serialized formats.
Finally, it is possible to index JSON fields in PostgreSQL databases, which expedites further queries on the unstructured data.

\begin{figure}
    \centering
    \includegraphics[width=\linewidth]{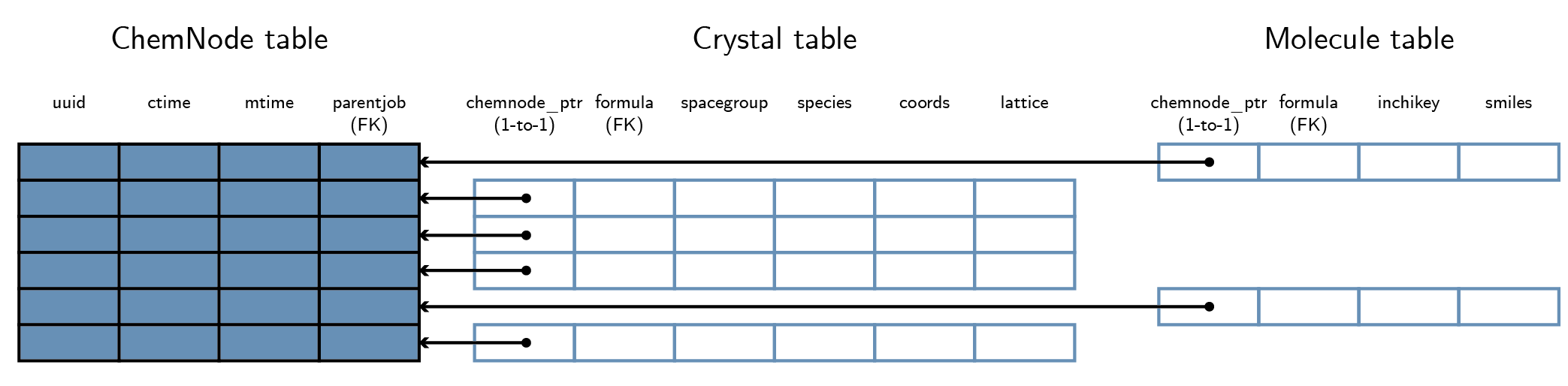}
    \caption{Tabular structure of ChemNodes and their relationship with inherited tables/models Crystal and Molecule.}
    \label{fig:table}
\end{figure}

The second option provides slightly more structure to the database, and is the default in this paper. 
In this approach, ChemNodes and CalcNodes are extended using multi-table inheritance in Django. 
This enables offspring models to automatically be considered a Node, but with additional structured data that describes the new models of interest.
Figure \ref{fig:table} showcases this approach with two models available by default in mkite, Crystals and Molecules. 
Crystals store information on crystal structures, i.e. systems containing an atomic motif and lattice parameters.
Molecules contain information on a chemical graph that represents a molecular system, usually encoded using SMILES strings and InChI hashes. 
A similar approach is used for CalcNodes, which are subclassed to originate tables containing energy/forces, DOS, and others.

Despite the additional overhead to the user/developer in creating and managing these database tables, benefits for their usage include standardization of data sharing/parsing across the database, jobs, and recipes.
Furthermore, although left joins are required when retrieving the ChemNodes or CalcNodes sub-tables, this approach provides an intuitive way to structuring diverse data while complying with data requirements for structuring workflows (see Sec. \ref{sec:workflow:struct}).

\section{Job/Workflow Creation}
\label{sec:workflow}

Once Jobs, ChemNodes, and CalcNodes are defined, workflows containing relationships between these structures can be created. 
Instead of requiring entire workflows to be added to the database as a whole unit, as often happens in FireWorks or AiiDA, mkite treats the creation of workflows as a synchronous, dynamic process.
This provides the user with more flexibility to alter calculation workflows as they are performed, filter segments of a workflow, and combine different branches of a workflow in downstream tasks.
Below, this workflow structure and creation is explained at length.

\subsection{Workflow structure}
\label{sec:workflow:struct}

\begin{figure}
    \centering
    \includegraphics[width=\linewidth]{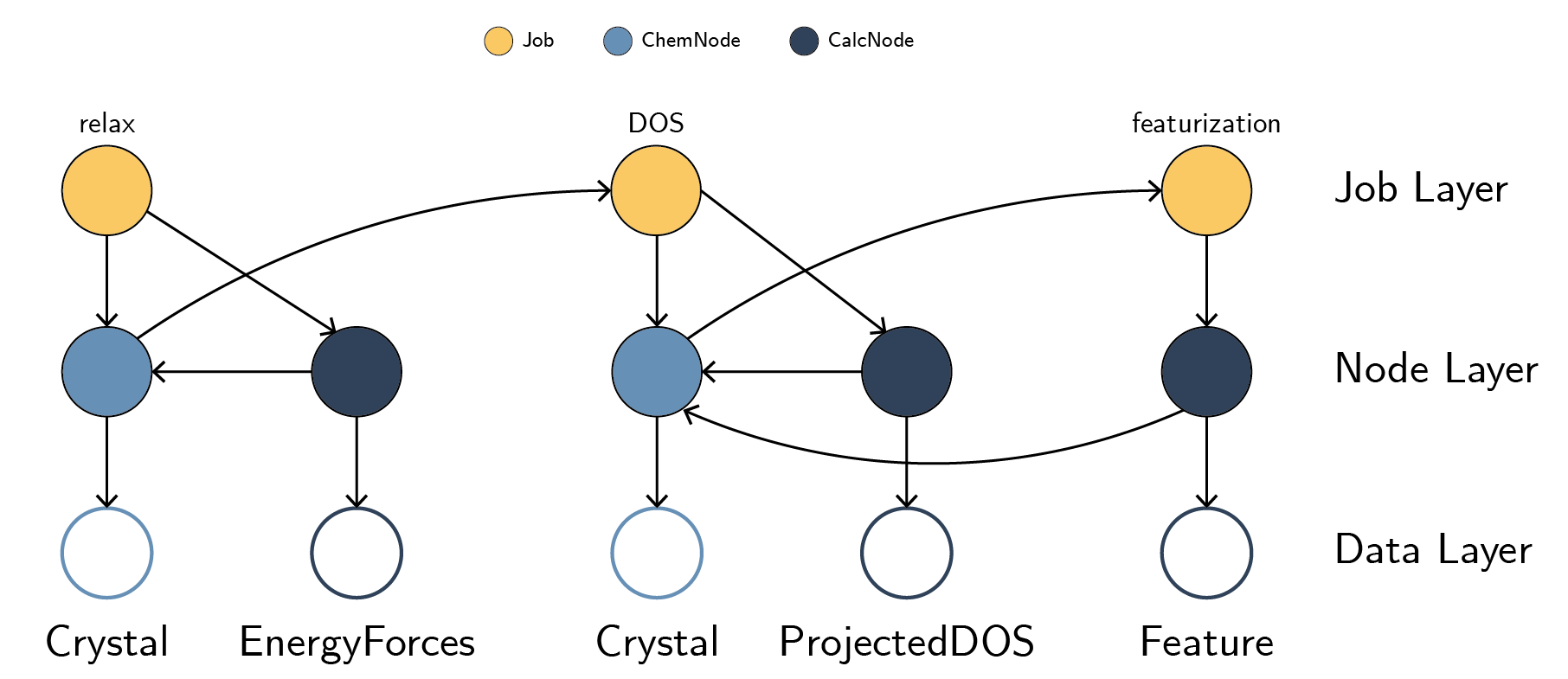}
    \caption{Diagram of a simple workflow implemented in mkite. Different Jobs are connected sequentially through ChemNodes. Each Node is usually associated to a data (empty circles), implemented as a table or JSON field (see Sec. \ref{sec:db:node}).}
    \label{fig:job-graph}
\end{figure}

As discussed previously, Jobs are central to the database schema, and connect input ChemNodes to output ChemNodes and/or CalcNodes. 
For example, a structure relaxation job takes a single structure as input, minimizes its energy/forces using a certain method, and outputs both a new structure and its corresponding energy/forces.
The user can also choose to save all relaxation steps along the way, in which case the Job has several ChemNodes and CalcNodes as output.
Once the relaxation is complete, the resulting ChemNodes can be used as inputs of new jobs. 
A different example is a featurization job, which may require one or more inputs, but outputs a single feature (CalcNode), which is then connected to the input ChemNode.
A workflow in mkite, therefore, is a series of Jobs connected by ChemNodes and whose outputs are ChemNodes and/or CalcNodes. 
Figure \ref{fig:job-graph} illustrates how this workflow is structured in mkite through database relations.

The relations between Jobs and Nodes shown in Fig. \ref{fig:job-graph} are enforced at the database level, and create three levels of abstraction: 
a Job layer, where a sequence of jobs can be visualized;
a Node layer, where the basic connections between ChemNodes, CalcNodes, and Jobs are created through database relations;
and a Data layer, which is associated to the data of each ChemNode or CalcNode, either in unstructured format in a JSON field or structured in a table (see Sec. \ref{sec:db:node}). 
As ChemNodes and CalcNodes can be created by only one Job, their relationship is defined through a foreign key, allowing a single Job to have multiple child nodes. 
On the other hand, one Job can have multiple inputs, as is the case of adsorption or docking algorithms, and the relationship between Jobs and their inputs is a many-to-many field (see Fig. \ref{fig:job-model}a for a visual representation of these relationships).

Within the constraints of a SQL database and Django, the Node layer is required to implement flexible workflows.
For example, traversing a workflow where different models are defined (e.g., those in Fig. \ref{fig:table}) may not be possible using SQL \texttt{JOIN} operations if the inverse relationship between a job and a ChemNode sub-table is not defined.
Instead, connecting Jobs to ChemNodes facilitates traversing a workflow by using the inverse relations \texttt{job == chemnode.parentjob $\leftrightarrow$ node $\in$ job.chemnodes} and \texttt{job $\in$ chemnode.childjobs $\leftrightarrow$ node $\in$ job.inputs} (Fig. \ref{fig:job-model}a).

\subsection{Job creation}
\label{sec:workflow:job}

Using the concepts defined above, workflows are defined by a sequence of jobs and inputs with relevant metadata. 
Creating workflows, therefore, corresponds to creating sequential rules for each job. 
Workflow managers such as AiiDA or FireWorks typically require these workflows to be defined beforehand, and then executed in a orderly fashion. 
This is similar to other approaches in more general workflow tools such as Airflow or Luigi, where entire workflow structures are  often defined beforehand and apply for all structures.

Nevertheless, in materials simulations, it is not uncommon to produce data in smaller batches and adapt the workflow after some analysis. 
Similar to experimental research, where different outcomes may require different procedures to be determined on-the-fly by the practitioner, automated simulations require flexibility in job creation. 
For example, a user may not require simulations at higher levels of theory for all enumerated systems in a database, but only for the ones with lowest energy among those in a low level of theory. 
Furthermore, in combinatorial spaces where different branches of calculations are performed independently, such as crystal-molecule interfaces, it is important to create jobs in a parallel manner.

To decouple job creation from the code and database, textual descriptions of the workflows are implemented in mkite. 
A job creation procedure requires specifying filters for input nodes, the options of the job, and the target recipe and experiment. 
Listing \ref{code:job-create} illustrates how to create a job using this textual description. 
The syntax of this specification matches a YAML file, which can be easily read both by humans and mkite commands, but it could have been written in equivalent formats such as JSON. 
In this example, Jobs of recipe name \texttt{vasp.pbe.relax} will be created for all inputs that match the given criteria:
(1) the ChemNodes pertain to experiment \texttt{01\_test}; 
(2) the ChemNodes were created from Jobs whose recipe is named \texttt{dbimport.MPImporter}; 
and (3) the ChemNodes are crystals whose formula contains the chemical element Au. 
The final job will be part of the experiment \texttt{01\_test} and will have the tag \texttt{bulk}, which are choices defined by the users to best organize their information. 
This is one of the major advantages of having Experiments and Projects as metadata of Jobs (see Fig. \ref{fig:job-model}).

\begin{code}
\captionof{listing}{YAML specification of job creation}
\label{code:job-create}
\begin{minted}{yaml}
# Simulation 1 of Experiment 01_test: relaxation
- out_experiment: 01_test
  out_recipe: vasp.pbe.relax
  inputs:
    - filter:
        parentjob__experiment__name: 01_test
        parentjob__recipe__name: dbimport.MPImporter
        crystal__formula__name__contains: Au
  tags:
    - bulk
\end{minted}
\end{code}

Using this input specification, Jobs can be created using one of mkite's commands, such as \texttt{create\_from\_file}. 
This command accesses a database, selects the inputs that do not have the target job (i.e., avoids creating duplicate tasks), and creates a Job model in the database connecting the inputs, options, experiment, and recipe. 
The jobs will be created with the status \texttt{READY}, indicating it has not been executed yet, but it has been created in the database (see Sec. \ref{sec:exec:build} for more details on Job status). 
The textual input should be saved by the user, desirably with version control, so that more jobs can be created in the future from the same pattern.
For example, if more Crystals are added to the database under the experiment \texttt{01\_test} and recipe \texttt{dbimport.MPImporter}, the same text input can be used to create new Job objects for the newly-added Crystals without adding duplicate information to the database. 
This is an important step towards automating complex workflows in materials simulations without hashing all structures.

An exception to the input specification above is the case of root jobs, which do not have any inputs. 
These jobs are used to add initial information to the mkite database from other sources, including files, literature, external datasets, and more. 
In Listing \ref{code:job-create} above, the job with recipe \texttt{dbimport.MPImporter} does not have any inputs other than a source file, and is considered a root job.
Similar commands to import files are provided in mkite, and can be further extended using the REST framework.

\subsection{Synchronous workflow creation}
\label{sec:workflow:create}

Using the framework defined above, workflows can be easily created by chaining textual descriptions of jobs. 
Continuing the example from Listing \ref{code:job-create},  additional specifications for other jobs can be created, as shown in Listing \ref{code:job-workflow}. 
The concatenation of these two Listings describes a workflow similar to the one in Fig. \ref{fig:job-graph}, where three jobs (relax, DOS, and feature) are executed in sequence. 

\begin{code}
\captionof{listing}{YAML specification of an mkite workflow}
\label{code:job-workflow}
\begin{minted}{yaml}
# Job 1 of Experiment 01_test: relaxation
- out_experiment: 01_test
  out_recipe: vasp.pbe.relax
  inputs:
    - filter:
        parentjob__experiment__name: 01_test
        parentjob__recipe__name: dbimport.MPImporter
        crystal__formula__name__contains: Au
  tags:
    - bulk

# Job 2 of Experiment 01_test: DOS
- out_experiment: 01_test
  out_recipe: vasp.pbe.pdos
  inputs:
    - filter:
        parentjob__experiment__name: 01_test
        parentjob__recipe__name: vasp.pbe.relax
  tags:
    - pdos

# Job 3 of Experiment 01_test: featurization
- out_experiment: 01_test
  out_recipe: ml.feature
  inputs:
    - filter:
        parentjob__experiment__name: 01_test
        parentjob__recipe__name: vasp.pbe.pdos
  tags:
    - feature
\end{minted}
\end{code}

By adding the jobs in this synchronous way --- that is, jobs are created only when a specific command is executed --- the user can control how often the database is accessed and queried for new data. 
Especially if coupled with a scheduling tool such as cron, the job creation can be performed automatically and periodically. 
Furthermore, this approach only creates downstream jobs for systems which have been successfully completed and filtered. 
This dynamic behavior allows changing the tasks and workflows ``on-the-fly'', according to the scope of the research, and easily attaching new jobs/workflow branches to existing data.

On the user side, adding new tasks to the computational workflow does not require accessing the database itself, but simply updating the workflow specification file, which is easily controlled, read, and revised. 
This approach to synchronous workflow creation is similar to a finite-state machine, where new steps are defined according to a ``clock'' and conditioned to the previous state. 
Especially with metadata such as experiment names, projects, and tags, jobs can be organized according to specific goals, facilitating querying the data and performing data analysis later. 
Finally, the synchronous job creation can be coupled with synchronous submission, which deploys the newly-created jobs for calculation outside of the local database.

\section{Job Submission/Execution}
\label{sec:exec}

After a job is created in the database, it can be submitted for execution in distributed hardware. 
However, one of the main design goals of mkite is the separation of concerns, decoupling job execution from the database. 
To enable job executions on the client side, a companion software package called mkwind manages submitted jobs without database access. 
In this section, the details of mkwind and job submission/execution will be detailed, along with the design criteria for distributing jobs across different computing environments.

\begin{figure}
    \centering
    \includegraphics[width=0.7\linewidth]{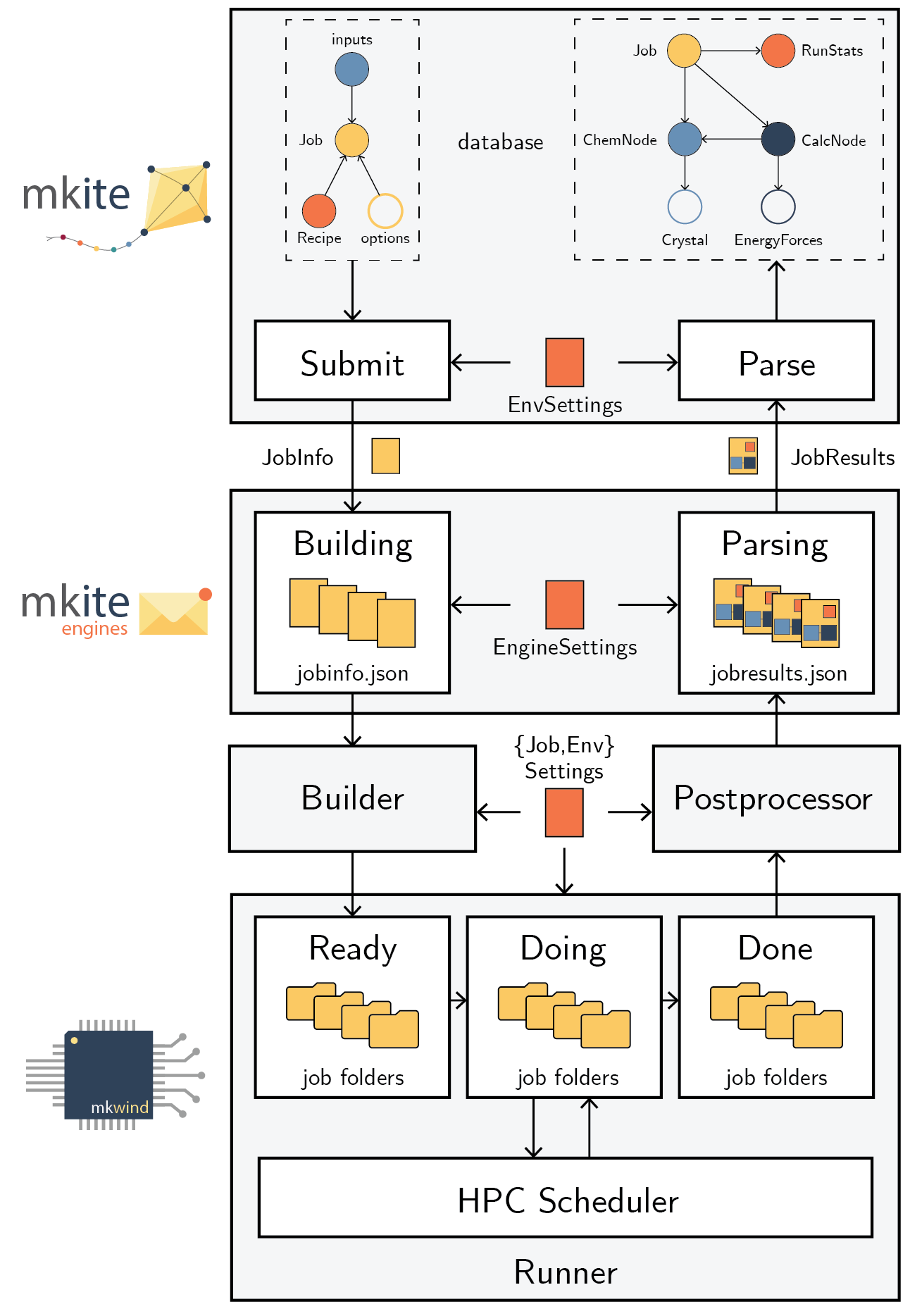}
    \caption{Overview of the coupling between mkite and mkwind. The workflow module of mkite handles Job database objects and submits their serialized representation (Fig. \ref{fig:job-model}b) to a message broker/queue. Then, remote mkwind clients read the message, build the job locally, interact with the HPC resources to execute the job, and postprocesses them before sending the results to mkite through the message queue. In the last step, the results of the job are deserialized and added to the database by mkite.}
    \label{fig:mkwind}
\end{figure}

As discussed before, mkite does not handles the job execution, but only submitting and parsing the serialized job metadata (Fig. \ref{fig:architecture}). 
On the other hand, mkwind is unaware of the existence of a database. 
The latter package simply executes Jobs based on input (serialized) data, processes the information on the local cluster, and returns a serialized file for consumption of mkite. 
In practice, this means that Jobs can be executed by mkwind in any computer without a database, allowing users to quickly submit calculations for systems that are not in the database itself.
This last feature enables prototype simulations to be performed, which is often useful in initial stages of simulations.
An overview of the coupling between mkite and mkwind is shown in Fig. \ref{fig:mkwind}. 

In the subsections below, the steps required to perform calculations, following Fig. \ref{fig:mkwind}, are explained. 
In particular, the role of mkwind and the communication protocols between mkite/mkwind explains how this system can be scaled to heterogeneous computing environments.

\subsection{Submitting and building jobs}
\label{sec:exec:build}

\begin{figure}
    \centering
    \includegraphics[width=0.7\linewidth]{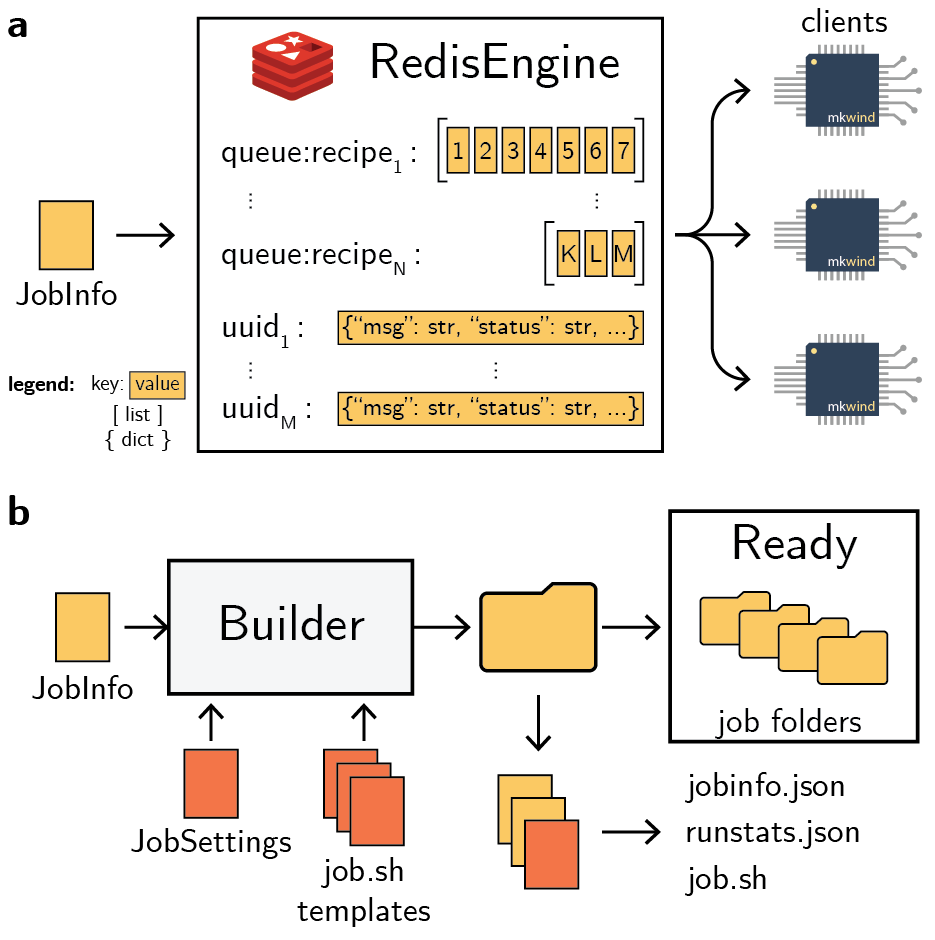}
    \caption{\textbf{a}, Overview of a Redis database used as a message broker, storing information about the jobs and scaling to numerous clients without concurrency issues. \textbf{b}, Once the information of a job is parsed by a client, the mkwind builder subroutine creates a folder containing the templates for submitting the job to an HPC scheduler.}
    \label{fig:redis}
\end{figure}

After the job creation procedure discussed in Sec. \ref{sec:workflow}, the database interfaced by mkite contains information on jobs ready for execution. 
During the submission stage, jobs from a selected recipe will be retrieved from the database, serialized, and submitted to a message queue. 
This submission can be performed using mkite's \texttt{submit} command, after which the status of the Job in the database will be changed to \texttt{RUNNING}, preventing it from being submitted again. 
The submission engine, shown in Fig. \ref{fig:mkwind}, uses a custom-made interface to a message broker/database such as Redis, as well as much simpler options such as a filesystem location where the JobInfo files will be saved prior to being built.
Figure \ref{fig:redis}a illustrates the workflow of an intermediate engine between mkite and mkwind in storing and distributing the jobs.
In this particular example, a Redis database is used as a message broker, storing temporary key-value pairs containing the serialized job information, and providing dedicated queues for jobs to be distributed according to their recipes.
The Redis cache can also be used to store temporary information on the job, such as in which client it is being executed, how many times it was restarted, cumulative execution time, and so on.
Despite the added complexity of a message broker, the ability to store and modify temporary information is an advantage in comparison with workflows containing only a production database.
For example, the cached information can be used to restart failed calculations from a checkpoint not stored in the production database.
In systems with a large number of atoms, this process can be automated to deploy sequential relaxations in different clients without directly transferring files.
Furthermore, the message broker further decouples the applications, speeds up the distribution of messages communication between clients, and prevent concurrency issues with atomic operations.

Once a job is submitted to the message broker, the production mkite database no longer controls its execution.
Instead, any instance of mkwind with access to the task queue prepares it for execution in the local HPC system. 
Nevertheless, given the architectural differences between HPC systems, the job must be tailor-built for the machine performing the calculation. 
For example, the desired number of cores for each recipe may vary from cluster to cluster, as well as the paths, modules, and execution scripts required to run each job. 
This is the role of the \texttt{build} command in mkwind. 
Instead of specifying this information at the server side or to use mostly on environmental variables, mkwind builds a job folder containing a script, the JobInfo file, and required specifications for running the task.
This enables providing different behaviors for each recipe in each HPC environment.
These specifications are controlled by a text file in the form of a YAML/JSON file which are loaded by mkwind once the build daemon is initialized.
Figure \ref{fig:redis}b illustrates the workflow of the job building process, which is controlled by a file such as the one shown in Listing \ref{code:build}.

\begin{code}
\captionof{listing}{YAML specification to build jobs using mkwind's \texttt{build} command}
\label{code:build}
\begin{minted}{yaml}
# Default specs for building jobs in a particular cluster
default:
  nodes: 1
  tasks_per_node: 18
  walltime: 24:00:00
  partition: pbatch
  account: test_account
  pre_cmd: |
    source $HOME/.jobrc
    source $HOME/mkite/bin/activate
  cmd: kite run
  post_cmd: |
    touch mkwind-complete

# Specs for VASP-specific recipes
vasp:
  tasks_per_node: 36
  account: production_account
  pre_cmd: |
    source $HOME/.jobrc
    source $HOME/envs/mkite/bin/activate
    module load vasp/6.3.1
\end{minted}
\end{code}

In Listing \ref{code:build}, a set of default settings is provided to mkwind to apply to all recipes (default).
For example, the CPU, partition, and walltime requirements are written explicitly in the configuration file (default, one node, 18 tasks/node, 24 hours of walltime). 
If different specifications are added to a particular recipe, the recipe settings override the default ones, allowing the user to fine-tune the behavior of each job for each HPC cluster. 
In this particular case, any recipe whose name starts with \texttt{vasp} creates a submission script that loads the VASP module in the HPC system prior to executing it, requests 36 tasks/node instead of the default 18 tasks/node, and charges the \texttt{production\_account} account instead of the default \texttt{test\_account}. 
Importantly, these specifications allow the user to determine partitions, queues, and accounts that the job will be executed in. 
This helps setting different HPC accounts for different projects, as well as selecting partitions according to the recipe to be executed. 
These records enable the creation of a folder that is served to the HPC job scheduler or the mkwind \texttt{runner} daemon, which will automate the process of executing the calculation.

As mentioned before, the mkite workflow attempts to decouple server/client processes for HTS.
Advantages of this decoupled behavior when submitting and building jobs include:
(1) the ability to manually add jobs to the task queue without a database;
(2) customization of the built jobs through hardware-specific scripts;
and (3) exemption from accessing a database at execution time.

\subsection{Running jobs}
\label{sec:exec:run}

After a job has been built, it can be executed manually by the user or submitted to job schedulers.
However, even for clusters containing schedulers such as SLURM, Torque, LSF, Flow, and others, the process of submitting and monitoring thousands of jobs or more can be tedious. 
To address this issue, mkwind provides a runner daemon, which monitors a folder called \texttt{ready} for new jobs to submit (Fig. \ref{fig:mkwind}). 
The daemon retrieves information from the scheduler to determine which jobs have been concluded, how many jobs are running, and how many are pending. 
Based on user-defined input, the daemon then decides whether to submit new folders for calculation or wait until new slots appear on the queue.
When a job is submitted, its folder is moved to the \texttt{doing} local folder. 
After the calculation is complete, the job is moved to a folder called \texttt{done}. 
These three folders are entirely managed by the mkwind runner daemon, as shown in Fig. \ref{fig:mkwind}.

\begin{figure}
    \centering
    \includegraphics[width=0.7\linewidth]{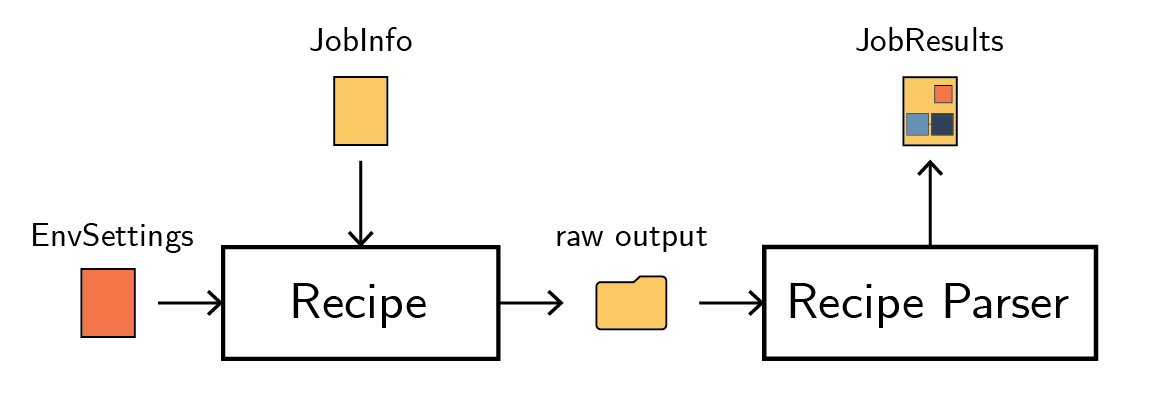}
    \caption{Overview of a recipe workflow at execution time. The recipe scripts produce the raw output from simulation software, and the parser converts it into the mkite serialized format.}
    \label{fig:recipe}
\end{figure}

On the job side, once a computational resource is allocated by the scheduler, the submitted script will run, loading modules and executing the task. 
If a recipe is executed from the \texttt{kite run} command, an mkite recipe will be loaded using the \texttt{recipe} module of mkite. 
Figure \ref{fig:recipe} shows how recipes are structured. 
The example recipe on Fig. \ref{fig:recipe} loads the JobInfo file, performs the task it was designed to do, and converts the raw, output files into a serialized file format.
In the process, the recipe may use external software by writing input files, calling the external software through an environmental variable or similar approach, and process the raw data produced by the other package. 
Compared to approaches where the results are added directly to a production database, the distributed parsing and serialization approach further optimizes the process of adding information to the database without overhead in the number of connections or concurrency issues.
This also enables other software packages to be used in conjunction with mkite or mkwind, given their decoupling.

One of the benefits of submitting jobs through a meta-scheduler is that the number of jobs pending or in execution can be controlled by the user in an automated fashion. 
Moreover, in sync with the builder daemon, the size of the \texttt{ready} folder can be controlled to a maximum of jobs waiting to be submitted. 
This combination of selective building/submission is one of the features that render mkwind an efficient software for job submission. 
Particularly in distributed computing environments, the availability of each computing resource is not known by the server unless additional scripts are added to interact with the main server.
In addition, HPC queues are dynamic and their utilization depends on factors such as number of users, fairshares, and so on. 
However, to maximize the computational utilization of potentially idle resources, users may want to distribute tasks to any available resource with the capability of executing them.
Therefore, clusters with fastest throughput will potentially have more slots for new jobs, which will be built and submitted according to their availability, as determined by mkwind.
On the other hand, clusters with lower computational throughput will have more jobs pending and less open slots, leading to slower consumption of incoming jobs from the message queue. 
Finally, mkwind's \texttt{runner} daemon is agnostic to the job added to the \texttt{ready} folder, and thus can be used to submit user-specific jobs that do not come from a database/build routine.

\subsection{Postprocessing and parsing jobs}
\label{sec:exec:postproc}

After the job was executed and moved to the \texttt{done} folder from mkwind's \texttt{runner} daemon, it is available for postprocessing. 
This consists in: 
(1) detecting whether the job finished successfully and produced results; 
(2) ensuring the results follow the schema for parsing in mkite; 
(3) sending the results to the message broker that will be accessible by mkite;
and (4) compressing and archiving the raw files produced by the job. 
The mkwind \texttt{postprocessor} daemon performs all these tasks as long as that adequate settings containing instructions to access the message queue and the \texttt{done} folder are provided.

Once the JobResults data is available in a messaging queue, mkite can retrieve this data using the \texttt{parse} command (Fig. \ref{fig:mkwind}). 
This procedure identifies the job that was submitted, deserializes the resulting ChemNodes, CalcNodes, and RunStats, and connects these nodes to the Job in the database. 
Finally, the status of the Job is changed to \texttt{DONE} to indicate no more action is required for this job. 
The data then becomes immediately available in the database, ready for data analysis or downstream tasks. 
The parsing step concludes the cycle of Job execution in the mkite framework.

\section{Extending and scaling mkite}
\label{sec:plugins}

In the descriptions above, the behavior of mkite and mkwind is conditioned to the database connections and models, the HPC environments, the availability of recipes, and so on. 
Moving from a high-level overview to a level closer to the execution of the framework requires understanding how the behavior of mkite/mkwind is controlled through text-based settings, and how to extend the basic functionalities of mkite to allow any recipe to be executed in any desired computational system. 
Below, these settings and possibilities for extension are detailed.

\subsection{Defining settings for heterogeneous environments}
\label{sec:plugins:configs}

\begin{figure}
    \centering
    \includegraphics[width=.5\linewidth]{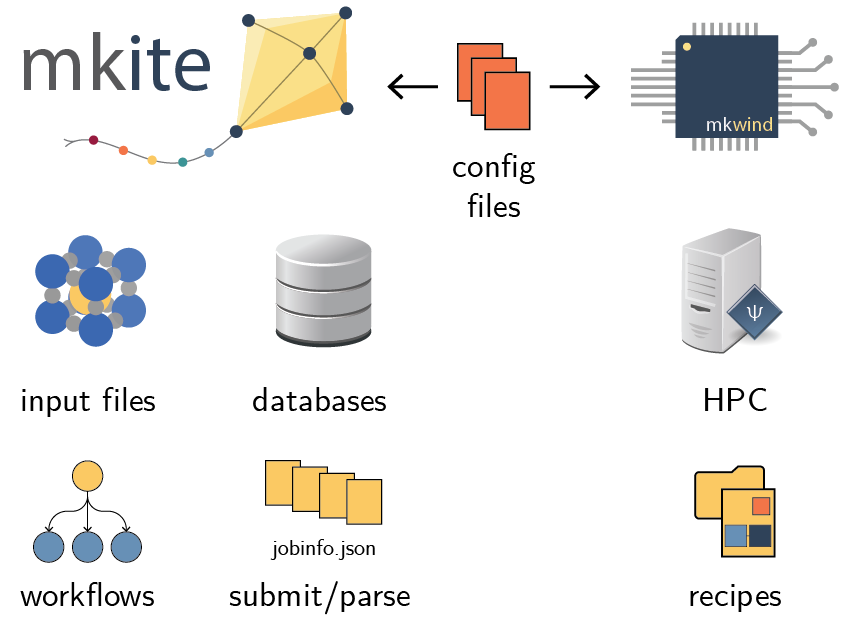}
    \caption{Example of configuration files that control the behavior of mkite and mkwind.}
    \label{fig:configs}
\end{figure}

In the subsections above, the role of heterogeneous environments, different queues, workflows, and databases were emphasized as fundamental part of mkite/mkwind. 
To achieve this diversity of configurations, a myriad of settings should be tailor-made for each environment, such as the recipe settings in Listing \ref{code:build}.
Example of files controlling the behavior of mkite/mkwind are shown in Fig. \ref{fig:configs}. 
These files hold the content of the Listings shown above, and are recommended to be under version control, as they define settings equivalent to ``dotfiles'' in Unix-like environments.

Configuring the behavior of mkite/mkwind, therefore, requires setting up configuration files for each HPC environment. 
For each one of these, the behavior of desired recipes can also be controlled at the mkwind building stage. 
For instance, settings for the mkwind runner change: what is the HPC scheduler in use, how many number of jobs are running, pending or waiting to be submitted, and more. 
Finally, options on message queues or databases available can be given to mkite to modify its behavior towards data submission/storage.
The collection of settings, workflows, input files, and scripts make mkite's behavior reproducible, amenable to version control, and easily distributed across systems of interest.

One example of setting for mkwind's runner is shown in Listing \ref{code:mkwind-settings}. 
Similarly to the other configuration/input files in this article, the description of the settings can be done with a YAML/JSON file. 
The number of jobs allowed to be running, pending, and ready are shown as \texttt{MAX\_RUNNING}, \texttt{MAX\_PENDING}, and \texttt{MAX\_READY}. 
This dictates the behavior of mkwind when submitting jobs to the scheduler specified at \texttt{SCHEDULER}. 
Templates for schedulers such as SLURM or LFS are already available in mkwind. 
Local jobs can be performed using pueue, a scheduler written in Rust for local machines, and whose template is available on mkwind.

\begin{code}
\captionof{listing}{YAML specification to setup mkwind}
\label{code:mkwind-settings}
\begin{minted}{yaml}
MAX_PENDING: 10
MAX_RUNNING: 100
MAX_READY: 10
SCHEDULER: pueue
BUILD_CONFIG: ${_self_}/../clusters/cluster1.yaml
ENGINE_EXTERNAL: ${_self_}/../engines/redis-cloud.yaml
ENGINE_LOCAL: ${_self_}/../engines/local.yaml
ENGINE_ARCHIVE: ${_self_}/../engines/archive.yaml
LOG_PATH: /home/user/logs
\end{minted}
\end{code}

Specifics of the settings will be kept up-to-date in the mkite/mkwind documentation,\footnote{\url{https://mkite.org}} along with all the settings required to submit the jobs.

\subsection{Extending mkite with new functionalities}

As in any automated simulations software, it is desirable to extend mkite beyond the built-in functionalities and recipes created by the lead developing team.
To enable calculations to be performed with any software, as well as integrating data generation, analysis, and featurization strategies with the database, mkite can connect to external Python packages that define an entry point to the \texttt{mkite} namespace.
This strategy, similar to AiiDA's plugin ecosystem, enables extending mkite beyond the currently supported recipes and packages. 
As these plugins decouple the development of the main mkite/mkwind softwares from the plugins, it can be customized in a per-user, per-project basis, providing new recipes, models, settings, and views for each user.
Furthermore, as different plugins can be installed in different Python environments (e.g., using Anaconda, virtualenv etc.), the types of simulations available for mkite can be customized for different projects simply by switching environments. 
In conjunction with potential environmental variables that can automate this approach, new tasks/behaviors for mkite can be personalized.
Tutorials on how to develop new recipes and plugins are available in the mkite documentation.

\section{Examples}
\label{sec:examples}

To exemplify how mkite can be used to automate workflows in materials simulation and discovery, three examples are described: (1) data extraction and analysis for the workflow in Fig. \ref{fig:job-graph}, (2) simulations of zeolite synthesis, and (3) simulations of surface catalysts. 
These examples can help consolidate the concepts described in the previous sections and their application to real materials systems.

\subsection{Example 1: Data Analysis}
\label{sec:analysis}

As the job workflow contains all relevant information and results to describe the calculations, analyzing mkite data usually starts with querying the results and parsing through them. 
In this context, the PostgreSQL database and Django ORM in the mkite package excel in filtering large volumes of data and parsing their results.
Furthermore, the metadata added to the jobs such as experiments and tags allow users to retrieve a specific set of results, or combine different datasets with informative labels, making the most of effective \texttt{JOIN}s in PostgreSQL databases.

Listing \ref{code:analysis} exemplifies how the data can be parsed from the database using Django models, assuming the connection with the database has been performed using a dedicated shell or through \texttt{django.setup()}. 
Built on the framework of Fig. \ref{fig:job-graph}, the query pre-selects the nodes of interest in a hypothetical mkite database and retrieves the relevant data from energies, DOS, and featurizations. 
In conjunction with Python packages such as Pandas and NumPy, concatenating the structural data, the energies data, and the ML features is a straightforward task, thus omitted from the example. 
As mkite uses Django as ORM, the queries can be performed simply by using Django's native syntax for queries, filtering, annotation, and more. 
If the user has access to the server where calculations are stored, this filtering and data retrieval can be performed in an interactive notebook such as a Jupyter server or an IPython shell, as well as automated in a Python script.

\begin{code}
\captionof{listing}{Python code to load the data from a database via Django}
\label{code:analysis}
\begin{minted}{python}
from mkite.orm import Job, Crystal

jobs = Job.objects.filter(
    experiment__name="01_test",
    recipe__name="vasp.pbe.pdos",
)
crystals = Crystal.objects.filter(parentjob__in=jobs)

data = crystals.values_list(
    "calcnodes__feature__value",
    "calcnodes__projecteddos__efermi",
    "parentjob__inputs__calcnodes__energyforces__energy"
)
\end{minted}
\end{code}

In Listing \ref{code:analysis}, Crystals generated by the \texttt{vasp.pbe.pdos} recipe are queried.
The \texttt{values\_list} method for the Django QuerySet is used to parse all the relevant information, with SQL \texttt{JOIN}s performed behind-the-scenes by Django.
In this particular example, the properties queried include those from parent nodes in the calculation graph, which can be accessed by the parent job's inputs.

One additional advantage of using Django and DRF is the ability to create views on top of the database. 
Django's built-in infrastructure for serving a web server can be used to automate the process of visualizing, aggregating, and analyzing the data in the mkite database. 
For example, a similar approach was used to create an interactive website for zeolite synthesis, where the data in a Django server can be retrieved, filtered, and analyzed by users without knowledge of the database structure \citep{Schwalbe-Koda2021a}.
This approach opens opportunities for extending mkite to provide views for specific models, plugins, and so on. 

\subsection{Example 2: application to zeolite synthesis}
\label{sec:example-1}

\begin{figure}
    \centering
    \includegraphics[width=\linewidth]{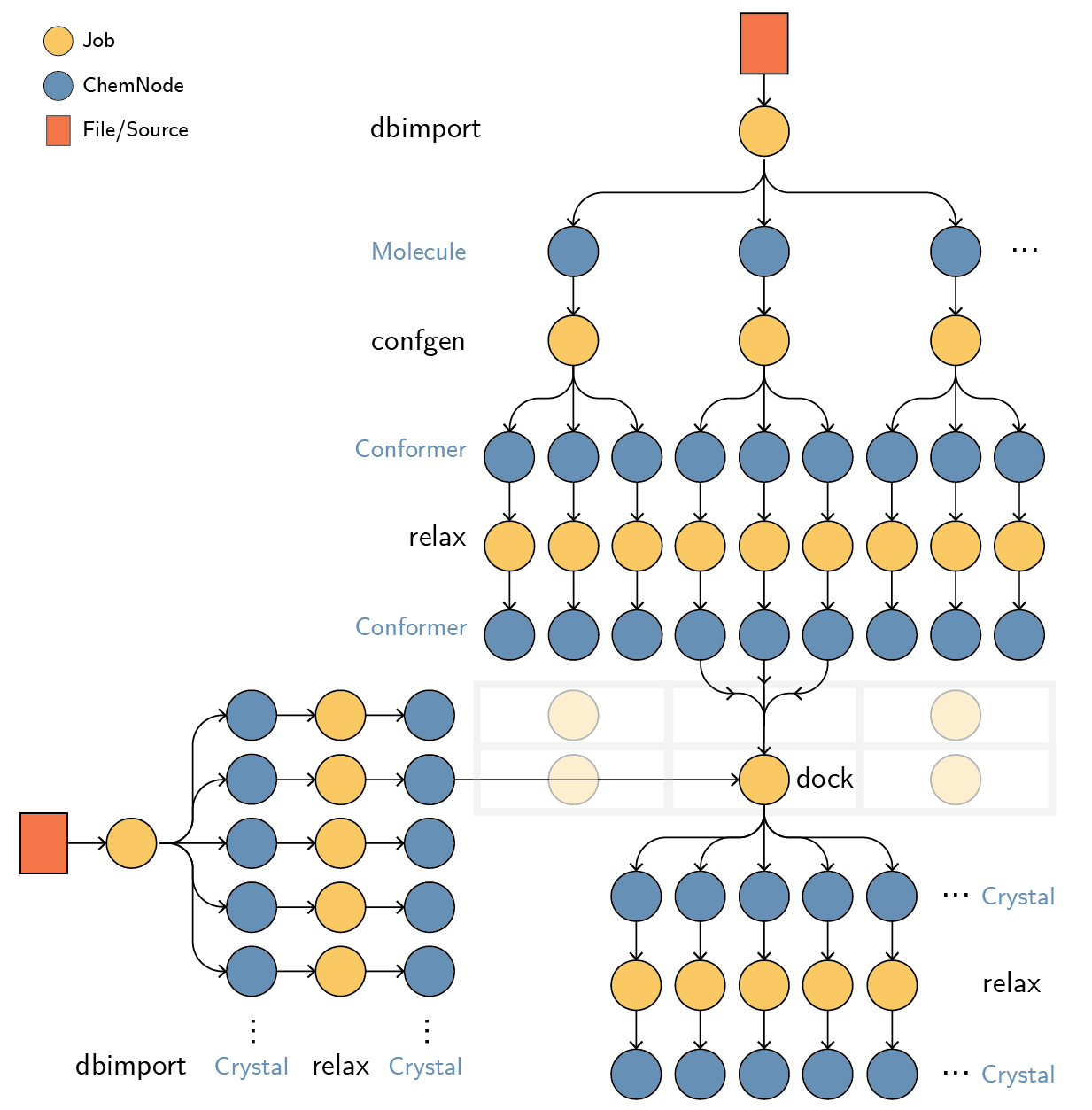}
    \caption{Example of workflow present in combinatorial simulation spaces of zeolite synthesis.}
    \label{fig:zeolite}
\end{figure}

An example where the space of simulations is combinatorial is that from zeolite synthesis.
In these materials, occlusion of organic templates in zeolite pores enables predicting the outcome topologies for zeolite synthesis from a computational perspective.
However, phase competition effects often play a role in which framework would be the outcome for a particular template.
The possible pairings between molecular conformers and zeolite frameworks is challenging to simulate from a single-branched workflow approach.
For example, the simulation requires several molecular conformers for each molecular graph, whose variable number of inputs depends on the molecular flexibility.
Furthermore, each docking attempt may generate various output structures, all of which can be analyzed.
Figure \ref{fig:zeolite} illustrates the steps involved in simulating interfaces of inorganic Crystals with organic molecular templates.

In Fig. \ref{fig:zeolite}, molecules and crystals are handled by different branches of the workflow, following the workflow of previous works \citep{Schwalbe-Koda2021Supramolecular,Schwalbe-Koda2021Benchmarking,Schwalbe-Koda2021a}.
As conformer generation/optimization can be computationally expensive for larger organic molecules, generating them for every calculation is not an efficient approach.
To handle these challenges, one can create workflows where calculations for molecules and crystals are treated as different Experiments.
Then, supramolecular interfaces can be created by taking molecular conformers and crystals as inputs for a docking algorithm.
As inputs are processed by recipes and not enforced at the database level, user-defined recipes can handle docking in several different approaches, including using multiple conformers and/or molecular species when performing docking.
This is useful in modeling higher-dimensional synthesis spaces in zeolites and beyond.

An example of the textual description of the workflow is available in \ref{sec:appendix:zeolite}.
Listings \ref{code:zeolite-workflow-1}, \ref{code:zeolite-workflow-2}, and \ref{code:zeolite-workflow-3} show how the two branches of the workflow can be performed independently, and connected using the Experiment metadata.
This is useful as the number of crystals and molecules increases substantially and calculations must be performed for both branches.
Listing \ref{code:zeolite-workflow-3}, in particular, shows how more than one input can be specified at the job creation stage (see also Sec. \ref{sec:workflow:job}).
As the inputs are not ordered and distinguishing between crystals and conformers can be performed at runtime, the database-blocking job creation process is fast, despite being combinatorial in nature.
The synchronous creation (Sec. \ref{sec:workflow:create}) also allows adding new molecules/crystals to the inputs and creating new jobs automatically, as each job is created as the new jobs are completed.

\subsection{Example 3: application to surface catalysts}
\label{sec:example-2}

In computational catalysis, automating simulations of adsorbate energies and reaction barriers enables probing combinatorial composition spaces. 
However, as the complexity of the reactions, surfaces, or composition grows, the computational steps involved in performing these calculations can become tightly coupled. 
For example, the GASPy software \citep{Tran2018ActiveEvolution,Tran2018DynamicScience} used for surface catalyst calculations uses Luigi as workflow manager to coordinate tasks such as adsorption or structure relaxation. 
Despite the success of this approach to small adsorbates, a different subspace of adsorbates, such as those from fine chemistry or biomass conversions, could increase the overhead for performing the tasks coupling the conformer generation and the subsequent adsorption.

\begin{figure}
    \centering
    \includegraphics[width=\linewidth]{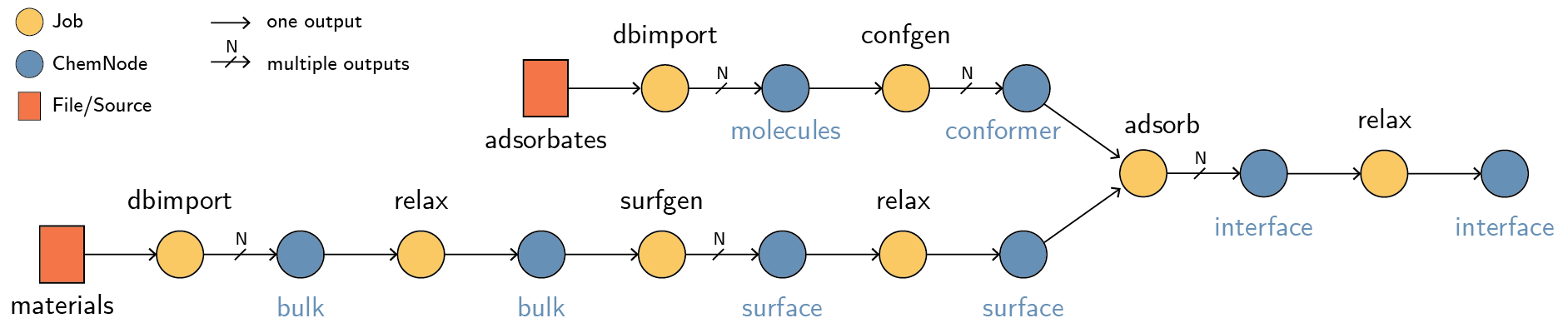}
    \caption{Example of workflow for simulating surface catalysts in mkite. Input crystals and adsorbates are created separately, and follow different branches of automation. Then, the \texttt{adsorb} task couples different surfaces and conformers, producing numerous interfaces ready to be simulated.}
    \label{fig:catalyst}
\end{figure}

Figure \ref{fig:catalyst} exemplifies a possible workflow for simulating combinatorial spaces for surface-adsorbate systems, as implemented in mkite. 
As the space of combinations is much larger than that of single-template zeolite synthesis in Sec. \ref{sec:example-1}, representing jobs from matrices of adsorbate-surface pairings is not easily achievable in two-dimensional workflows.
To handle this visualization, jobs that generate multiple outputs are represented by an arrow with a cross and an ``N'', and jobs generating single inputs are represented by an uncrossed arrow.
Starting from a list of adsorbates and bulk materials of interest, the workflow relaxes the bulk structures, creates the surfaces, relaxes surfaces, creates conformers for the molecules of interest, adsorbs the conformers onto surfaces, and relaxes the interfacial systems.
Decoupling these tasks enables recipes such as \texttt{confgen} to be used in other workflows, while still providing agility for prototyping the workflow.
For example, using the job creation scheme described in Sec. \ref{sec:workflow}, one can quickly assemble the sequence of steps shown in Fig. \ref{fig:catalyst}, following Sec. \ref{sec:example-1}.
An example of a possible textual description for this workflow, following the notation from Listing \ref{code:job-workflow}, is available in \ref{sec:appendix:catalyst}.

As is the case of combinatorial spaces, as the workflow produces a large number of input structures such as different surfaces and different materials, the space of simulations rapidly grows. 
Nevertheless, creating jobs for the simulations is straightforward using the description shown in Listings \ref{code:catalyst-workflow-1}, \ref{code:catalyst-workflow-2}, and \ref{code:catalyst-workflow-3}.
Often, however, not all structures have to be simulated.
For example, the workflow can be controlled by manually adding filters to Listing \ref{code:catalyst-workflow-3}, selecting which adsorptions will be performed.
This restriction still enables a larger chemical space to be explored later, without having to discard the information that has already been created in the database.
Furthermore, control of job execution may be possible by using tags, which enable the user to select which jobs to submit.
Finally, active learning strategies can be easily integrated to the workflow by adding new structures into the pipeline, often simply by updating the input file/source or adding a new one.
The synchronous job generation approach automates the process of simulating catalyst structures without tasks tightly coupled to a task manager, improving the flexibility of the process for the developer/researcher.

\section{Discussion}
\label{sec:disc}

\subsection{Comparison with existing software}
\label{sec:disc:comparison}

As automated, HTS approaches become more popular, so do the engines enabling their workflow. 
As mentioned in the Introduction, a variety of workflow engines are available at the time of writing of this paper. 
While more mature softwares typically exhibit more features, their approaches to enable HTS are distinct to that of mkite. 
Below, a few design choices exemplify the niche of mkite in HTS software and how it complements other systems for data management and job submission in atomistic simulations.

\textbf{AiiDA} \citep{Pizzi2016,Huber2020AiiDAProvenance,Uhrin2021WorkflowsWorkflows} is the most similar approach to mkite. 
Its mature code, plugins, and community have enabled the application of this software in a myriad of simulations. 
Several choices adapted in mkite build on successes proposed by AiiDA, such as the ideas of plugins to extend database models or recipes. 
Differently from AiiDA, however, mkite decouples job submission from execution. 
This means that the data provenance is not extensively recorded in the database as performed in AiiDA, as mkite recipes process the data independently without storing all information.
Although this lowers the fidelity of the data provenance graph in mkite, it also enables a higher flexibility of mkite in distributing the calculations and job decoupling.
Another useful feature of mkite is the ability to use Django's native language to perform queries, data analysis, and produce views. 
This means that features from the Django community to set up front-ends are readily available for mkite. 

\textbf{FireWorks} \citep{Jain2015Fireworks:Applications} is another popular software for submitting complex workflows in HPC environments. 
FireWorks workers can be setup in any environment and interact with schedulers as the data becomes available.
It also has powerful error handling engines, stores job metadata in a MongoDB database, and allows distributing jobs in various computing environments.
This enables concurrent processing of job information and submission.
Although the FireWorks software has enabled the construction of widely successful databases, including the Materials Project, their use can be challenging when workflows with more than one branch are required.
The main difference between mkite and FireWorks is the ability to connect different workflow branches at the database level, as shown in Sec. \ref{sec:workflow} of this paper.
This is particularly useful when concatenating jobs requiring different software packages.
Tailoring job building in mkite can also allow greater flexibility in job submission, for example, when selecting different accounts to be charged depending on the projects and queues.
Finally, mkwind's functionality of executing jobs without requiring a database can facilitate prototyping tasks compared to FireWorks' robust database infrastructure.

\textbf{pyiron} \citep{Janssen2019Pyiron:Science} and \textbf{signac} \citep{Adorf2018SimpleFramework} are more recent computational tools for materials simulation showing wide usage.
Building on the niche of storing larger trajectories as opposed to smaller jobs, pyiron and signac offer an IDE for performing jobs and distributing them on HPC resources.
This includes classes for storing projects, results, connecting jobs, visualizing the data and more, striving to be an ``electronic lab notebook'' for computational materials scientists.
Whereas mkite also shares this philosophy of facilitating computational experiments, it focuses less on building an integrated IDE and more on job submission/distribution tools for combinatorial spaces.
For example, mkite allows jobs to be filtered/created with text-based inputs in combinatorial spaces, bypassing the need of defining workflows using Python.
This becomes particularly useful as larger search spaces are explored.
Another feature of mkite is the ability to submit jobs in a client-agnostic way, thus bypassing the need for defining job specifications on the server side, as usually performed in pyiron or signac.
Whereas the latter packages can still be used seamlessly with schedulers and queues, their coupling with the data storage may become more challenging as the number of client workers increases.

Python-based packages such as \textbf{atomate(2)} \citep{Mathew2017Atomate:Workflows}, \textbf{quacc} \citep{Rosen2019IdentifyingTheory}, \textbf{molSimplify} \citep{Ioannidis2016molSimplify}, and many others are complementary to mkite in functionality. 
These tools facilitate job creation, execution, and parsing through well-developed scripts produced in Python, abstracting away from input files and software calls.
Whereas some of these tools, such as atomate, enable saving results directly to a database, their use is not restricted to a specific data target.
On the other hand, mkite and mkwind provide tools for keeping a record of jobs, concatenating workflows, and managing jobs in distributed computing environments.
As such, these packages aiding job execution can be used in mkite recipes and collaborate with mkite's workflow orchestration and data storage.

Several other community packages such as the \textbf{Atomic Simulation Environment} (ASE) \citep{HjorthLarsen2017}, \textbf{pymatgen} \citep{Ong2013}, and others can be coupled with mkite to provide data representation, storage, and workflow orchestration.
For example, they can provide a schema for the data when multi-table inheritance is not used (see Sec. \ref{sec:db:node}) or be used directly in recipe execution.
The comprehensive processing abilities and interfaces to several software packages, as also done by \textbf{cclib} \citep{O'boyle2008cclib} and similar softwares, can expedite the creation of mkite recipes for job execution.

\subsection{Limitations and future extensions}
\label{sec:disc:limits}

Preliminary results of mkite have shown that mkwind is able to scale to thousands of jobs running in parallel on nodes with different architectures. This can increase the throughput of simulations by submitting embarassingly parallel jobs to efficient computing resources and handling the data integration in an agnostic manner. However, the current framework still requires the user to setup the environment prior to running the calculations. For example, the Python environment used to run the mkwind daemons have to be set up in each HPC system, along with the relevant configuration files for the recipes of interest. Whereas this is somewhat unavoidable in large, production systems, it imposes a limitation for scaling this platform on local, volunteer computing resources. In this case, a larger integration with existing software has to be pursued to scale job execution ability and recipes in global computing networks.

This perspective also requires scaling up message distribution software. Current queues are based on shared filesystem folders and on Redis. Although the latter can scale up to tens of thousands of simultaneous connections and millions of operations per seconds, transmitting information of large calculations, such as density of states of systems with hundreds of atoms, can be challenging with these protocols, even within serialized data. This is a benefit of directly writing results into a database after postprocessing, although that comes at the cost of blocking connections to the production database itself.

As the mkite package integrates with Django, visualizations for the database can be implemented on the user side, enabling servers to exhibit interactive data.
Although interactive views/filters can build on existing models and serializers, no default visualizations are provided by mkite.
This is a direction that will be explored in the future, especially for retrieving job statistics and visualization of simulated results.

Finally, recipes in mkite do not enforce input formats or processing restrictions.
This added flexibility enables users to propagate arbitrary information through the workflow chain, storing them as unstructured data in ChemNodes and CalcNodes.
Nevertheless, this feature can also impact job execution if custom information is lost during the use of an unstructured schema.
Thus, users should be responsible for keeping enough data to enable downstream tasks to be performed adequately.

\section{Conclusion}
\label{sec:concl}

The mkite toolkit is a high-throughput simulation software tailored for multi-input, multi-environment simulations. 
Built using the server-client design pattern, the mkite package decouples database access from simulation tools, enabling any configured client to perform calculations as soon as they are available.
Furthermore, through synchronous database updates and text-based workflow descriptions, users can create and modify complex workflow structures without altering the database structure.
This is particularly useful when exploring combinatorial materials spaces such as interfacial properties, supramolecular systems, or materials discovery approaches.
Finally, mkite recipes can be easily connected to existing packages, simplifying the process of creating new tasks or handling input/output files from external software.
The design principles in mkite carve a niche among other high-throughput simulation packages and facilitate the development of databases, workflows, and front-end tools using PostgreSQL, Django, and Python.

\section{Acknowledgements}
\label{sec:ack}

This work was performed under the auspices of the U.S. Department of Energy by Lawrence Livermore National Laboratory under Contract DE-AC52-07NA27344, funded by the Laboratory Directed Research and Development Program at LLNL under project tracking code 22-ERD-055. The author acknowledges Vincenzo Lordi, Brandon Wood, and Tuan Anh Pham for conversations regarding this project, and LLNL's Workflow Enablement Group for help with internal software.

Manuscript released as \texttt{LLNL-JRNL-843647}.

\section{Code Availability}
\label{sec:code}

The mkite suite is released under the Apache 2.0 (with LLVM exception) license. The code is available at \url{https://github.com/mkite-group/}. The user guide is available at \url{https://mkite.org}. Release code: LLNL-CODE-848161.

\appendix

\section{Workflow scripts for Example 1}
\label{sec:appendix:zeolite}

\begin{code}
\captionof{listing}{Workflow for Crystal branch of zeolite workflow in Fig. \ref{fig:zeolite}.}
\label{code:zeolite-workflow-1}
\begin{minted}{yaml}
# Job 1 of Crystal branch
- out_experiment: 01_crystals
  out_recipe: gulp.slc.relax
  inputs:
    - filter:
        parentjob__experiment__name: 01_crystals
        parentjob__recipe__name: dbimport.CrystalFileImporter
  tags:
    - unloaded
\end{minted}
\end{code}

\begin{code}
\captionof{listing}{Workflow for Molecule branch of zeolite workflow in Fig. \ref{fig:zeolite}.}
\label{code:zeolite-workflow-2}
\begin{minted}{yaml}
# Job 1 of Molecule branch
- out_experiment: 02_molecules
  out_recipe: conformer.generation
  inputs:
    - filter:
        parentjob__experiment__name: 02_molecules
        parentjob__recipe__name: dbimport.MolFileImporter
  tags:
    - conformer
    
# Job 2 of Molecule branch
- out_experiment: 02_molecules
  out_recipe: gulp.dreiding.relax
  inputs:
    - filter:
        parentjob__experiment__name: 02_molecules
        parentjob__recipe__name: conformer.generation
        parentjob__tags__name: conformer
  tags:
    - conformer
\end{minted}
\end{code}

\begin{code}
\captionof{listing}{Workflow for Crystal-Molecule branch of zeolite workflow in Fig. \ref{fig:zeolite}.}
\label{code:zeolite-workflow-3}
\begin{minted}{yaml}
# Job 1 of Joint branch
- out_experiment: 03_joint
  out_recipe: void.dock
  inputs:
    - filter:
        parentjob__experiment__name: 01_crystals
        parentjob__recipe__name: gulp.slc.relax
    - filter:
        parentjob__experiment__name: 02_molecules
        parentjob__recipe__name: gulp.dreiding.relax
  tags:
    - joint
    
# Job 2 of Joint branch
- out_experiment: 03_joint
  out_recipe: gulp.dreiding.relax
  inputs:
    - filter:
        parentjob__experiment__name: 03_joint
        parentjob__recipe__name: void.dock
  tags:
    - joint
\end{minted}
\end{code}

\section{Workflow scripts for Example 2}
\label{sec:appendix:catalyst}

\begin{code}
\captionof{listing}{Workflow for Crystal branch of surface catalyst workflow in Fig. \ref{fig:catalyst}.}
\label{code:catalyst-workflow-1}
\begin{minted}{yaml}
# Job 1 of Crystal branch
- out_experiment: 01_crystals
  out_recipe: vasp.rpbe.relax
  inputs:
    - filter:
        parentjob__experiment__name: 01_crystals
        parentjob__recipe__name: dbimport.CrystalFileImporter
  tags:
    - bulk
    
# Job 2 of Crystal branch
- out_experiment: 01_crystals
  out_recipe: catalysis.surfgen
  inputs:
    - filter:
        parentjob__experiment__name: 01_crystals
        parentjob__recipe__name: vasp.rpbe.relax
        parentjob__tags__name: bulk
  tags:
    - surface
  options:
    max_miller_index: 2
    min_vacuum_size: 15
    
# Job 3 of Crystal branch
- out_experiment: 01_crystals
  out_recipe: vasp.rpbe.relax
  inputs:
    - filter:
        parentjob__experiment__name: 01_crystals
        parentjob__recipe__name: catalysis.surfgen
        parentjob__tags__name: surface
  tags:
    - surface
\end{minted}
\end{code}

\begin{code}
\captionof{listing}{Workflow for Molecule branch of the surface catalyst workflow in Fig. \ref{fig:catalyst}.}
\label{code:catalyst-workflow-2}
\begin{minted}{yaml}
# Job 1 of Molecule branch
- out_experiment: 02_molecules
  out_recipe: conformer.generation
  inputs:
    - filter:
        parentjob__experiment__name: 02_molecules
        parentjob__recipe__name: dbimport.MolFileImporter
  tags:
    - conformer
\end{minted}
\end{code}

\begin{code}
\captionof{listing}{Workflow for Crystal-Molecule branch of the surface catalyst workflow in Fig. \ref{fig:catalyst}.}
\label{code:catalyst-workflow-3}
\begin{minted}{yaml}
# Job 1 of Joint branch
# The filter selects post-relaxation Crystals with tag surface
- out_experiment: 03_joint
  out_recipe: catalysis.adsorption
  inputs:
    - filter:
        parentjob__experiment__name: 01_crystals
        parentjob__recipe__name: vasp.rpbe.relax
        parentjob__tags__name: surface
    - filter:
        parentjob__experiment__name: 02_molecules
        parentjob__recipe__name: conformer.generation
  tags:
    - interface
    
# Job 2 of Joint branch
- out_experiment: 03_joint
  out_recipe: vasp.rpbe.relax
  inputs:
    - filter:
        parentjob__experiment__name: 03_joint
        parentjob__recipe__name: catalysis.adsorption
        parentjob__recipe__tags: interface
  tags:
    - interface
\end{minted}
\end{code}

\end{document}